\documentclass[preprint,amsmath,amssymb,floatfix]{revtex4}
\usepackage{graphicx}
\usepackage{dcolumn}
\usepackage{bm}
\begin{document}

\begin{center}
{\large \bf Macro- and Micro-structure of Trust Networks}
\end{center}

A challenging problem in the social sciences is the characterization of the
formation of ``social capital'' \cite{Wasserman94,Putnam}.  It is believed
that societies with more social capital are more democratic and economically
developed than societies with little social capital \cite{Putnam}.  However,
social capital is a concept hard to quantify and measure in a ``real-word''
context.  Here, we take advantage of an existing ``web of trust'' between
users of the "Pretty-Good-Privacy" (PGP) encryption algorithm
\cite{Stallings95,Garfinkel94}.  The PGP algorithm is used in digital
communication to ``sign'' documents so that the recipient knows for sure who
the author is.  Our analysis reveals the coexistence in the web of trust of a
macro scale-free structure with micro strongly-connected cells in a complex
network \cite{Watts98,Amaral00,Strogatz01,Albert02}.  We also show that when
this network is intentionally attacked \cite{Albert00} the scale-free
structure \cite{Barabasi99} rapidly disintegrates and that the resulting
network is partitioned into a large number of small strongly-connected cells
that are resilient to the intentional attack.

In digital communication, as in clandestine organizations, there is the
distinct possibility that identities will be forged.  Hence, there is the
need to develop systems by which one can trust that someone is who he claims
to be.  A possible solution is a centralized service that certifies users;
another is a self-organized system in which users certify one another.  In
the PGP web of trust, a user certifies another by ``signing'' his public
encryption key \cite{Stallings95,Garfinkel94}.  For such a decentralized
solution to work, the system has to build social capital---i.e., to add more
signatures---otherwise it will not work efficiently.  Hence, we suggest that
the PGP web of trust is a social network for which one can {\it measure\/}
the formation of social capital and also the structure of the network itself
(Fig.~1a).

We calculate the in- and out-degree cumulative distributions for the
web of trust, and find that this distribution decays as a power law
$P(k) \propto k^{-\alpha} \,$, with exponents $\alpha_{in} = 1.8 \pm
0.2$ and $\alpha_{out} = 1.7 \pm 0.2$ (Fig.~1b).  We also find that
the network is not a connected graph but instead comprises many
strongly-connected clusters with a very wide range of sizes (from 2 to
approximately 10,000 nodes).  This feature of the network enables us
to study its structure in more detail by considering the properties of
each cluster separately: We measure the clustering coefficient
$C$---which quantifies to which extend nodes adjacent to a given node
are also adjacent to each other \cite{Watts98}---for each of the
clusters in the network and find that it remains approximately
constant (Fig.~1b).  This result is rather surprising since $C$
decreases rapidly with cluster size for a ``pure'' scale-free network
\cite{Barabasi99} or a random graph \cite{Watts98}.

The strongly-connected clusters are cross-linked by means of a few
high degree nodes that act as ``hubs'' (Fig.~1c).  The hubs appear to
be organized in a hierarchical fashion giving rise to the scale-free
structure while the strongly-connected clusters give rise to the large
value of $C$.  The coexistence of the scale-free connectivity and the
strongly-connected clusters has implications for the vulnerability of
trust networks. To demonstrate this assertion, we apply to the largest
cluster the attack method proposed in Ref.~\cite{Albert00}.  As shown
in Fig. 1d, we find that: (i) The network is {\it extremely fragile\/}
against the systematic removal of the most connected nodes---due to
the highly skewed connectivity of the hub super-structure; and (ii)
the ``lower-level'' strongly-connected structure remains essentially
un-affected. These two consequences of the structure of the web of
trust have implications regarding the potential recovery of the
network after an attack targeting the most connected nodes.  Indeed,
because the network still comprises large strongly-connected cells
even after the largest cluster has been broken, it would be possible
to quickly rebuild a large fraction of the original network by
creating just a few links.

An important point that does not escape our attention is that there are
plausible similarities between the structure of the web of trust we analyze
here and the structure of clandestine/secret organizations \cite{Krebs02}.
In clandestine organizations trust is an essential ingredient because each
member has has to be confident that her associates are not going to betray
her.  Moreover, as we found for the web of trust, clandestine organizations
comprise small strongly-connected cells which are then connected to one
another through a few highly-connected individuals \cite{Krebs02}.


\vfill
\noindent
{\bf X. Guardiola$^1$, R. Guimer\`a$^2$, A. Arenas$^3$,
A. Diaz-Guilera$^{1,2}$, D. Streib$^4$, L. A. N. Amaral$^5$}

\noindent
{$^1$ Departament de F\'{\i}sica Fonamental, Universitat de
Barcelona, 08028 Barcelona, Spain \\
$^2$ Departament d'Enginyeria Qu\'{\i}mica, Universitat Rovira i
Virgili, 43006 Tarragona, Spain \\
$^3$ Departament d'Enginyeria Inform\`atica i Matem\`atiques,
Universitat Rovira i Virgili, 43006 Tarragona, Spain \\
$^4$ Free Standards Group \\
$^5$ Center for Polymer Studies and Department of Physics, Boston
University, MA 02215, USA \\
}

\newpage


\begin{figure}
 \caption{ Growth and structure of the PGP web of trust
  \protect\cite{Garfinkel94}. The links in the PGP data set refer to the
  signatures of users public keys by other users.
%
  {\bf a}, 
  Number of public keys in two PGP databases
  (``ftp://ftp.uit.no/pub/crypto/pgp/keys/pubring.pgp'' and
  ``http://dtype.org'') and average path length between those keys.  We find
  that the number of signatures increases exponentially while the average
  path length increases linearly.  These results can be interpreted as an
  increase in the social capital of the PGP web of trust because the number
  of users---i.e., of public keys---connected efficiently---i.e., only a few
  degrees of separation apart---is increasingly rapidly with time.
  {\bf b}, Cumulative distribution of in- and out-degrees for the PGP web of
  trust.  We analyze data recorded on July 2001 at ``http://dtype.org'', when
  the web comprised 191,548 individuals and 286,290 directed links between
  them.  The data follow a power law dependency with exponents -1.8 for the
  in-degree and -1.7 for the out-degree.
  (Inset) Clustering coefficient $C$ for the different strongly-connected
  clusters. For comparison, we also show the values of $C$ for (i) the
  small-world network model of \protect\cite{Watts98}, (ii) the scale-free
  network model of \protect\cite{Barabasi99}, and (iii) a random graph
  \protect\cite{Watts98}. $C$ is approximately constant for the web of trust,
  while for (ii) and (iii) it decays rapidly with cluster size.
  {\bf c}, A strongly-connected cluster comprising 21 nodes.  This cluster is
  strongly connected because every node is reachable from any other
  node. White lines indicate bi-directional links while yellow arrows
  indicate uni-directional links. The red nodes indicate the two
  strongly-connected clusters that would be left if one removes the green
  node (the hub).
  {\bf d}, Intentional attack on the nodes with the highest in-degree
  of (i) the largest strongly-connected cluster of the web of trust and (ii)
  a random graph with the same in- and out-degree distributions.  Initially,
  both graphs have 9562 nodes with an average degree of 5.80. As the fraction
  $f$ of nodes removed increases, the cluster is split into smaller
  components: (Top) Relative size $S$ of the largest strongly-connected
  cluster. (Bottom) Average size $\langle s\rangle$ of the other
  strongly-connected clusters.  Note that for the web of trust the largest
  strongly-connected cluster breaks down faster but that the other
  strongly-connected clusters have average sizes that remain unchanged up to
  the total destruction of the largest strongly-connected cluster. For the
  random graph, the small clusters formed by removing nodes are almost all
  isolated nodes, which explains the slower decrease of $S$ and also the
  constant value of $\langle s\rangle=1$.}
 \label{figure1}
\end{figure}

\begin{figure}
\centerline{
 \includegraphics*[width=0.45\columnwidth]{social1}\quad
 \includegraphics*[width=0.45\columnwidth]{degree2}}

 \vspace*{0.5cm}
 \centerline{
 \includegraphics*[height=0.40\columnwidth]{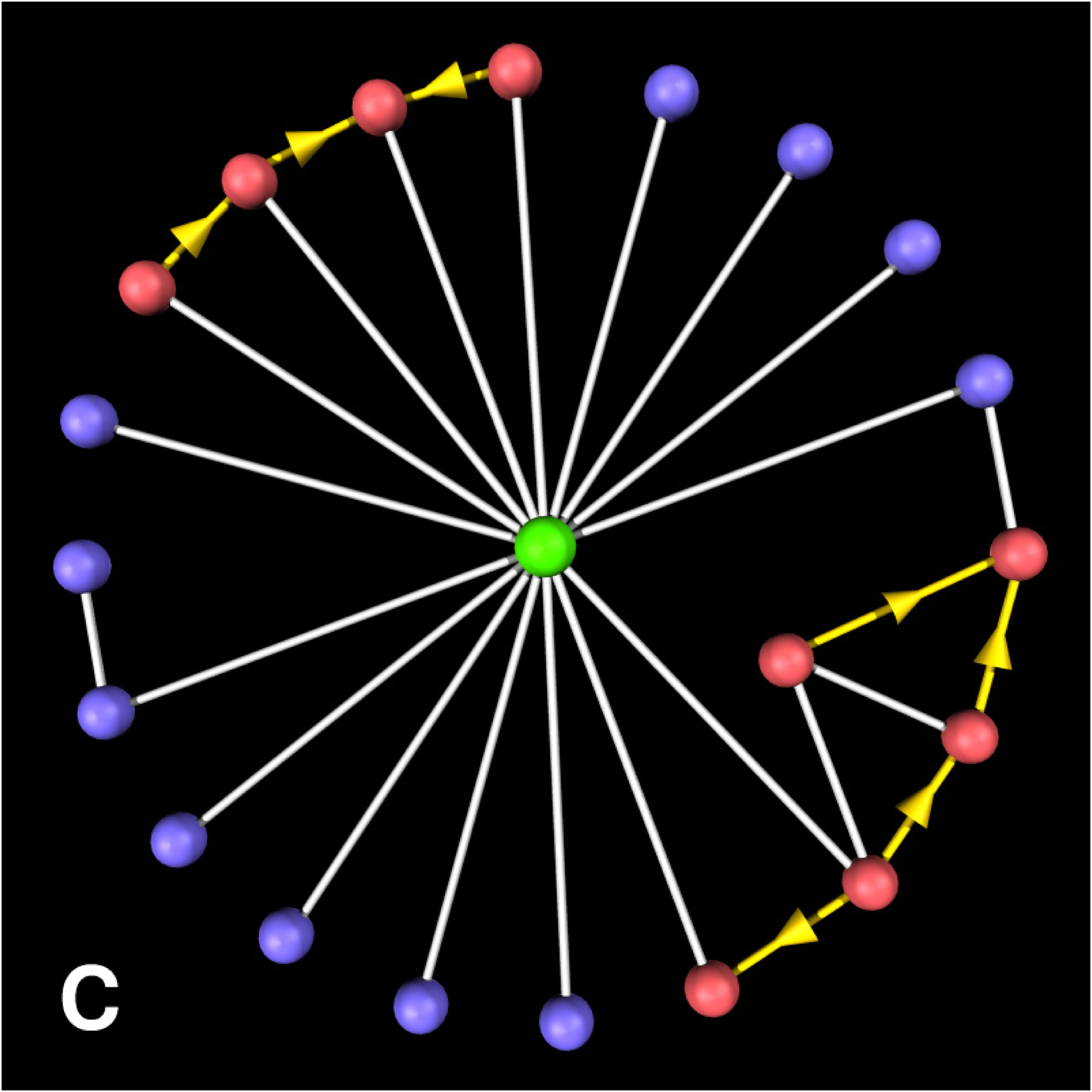}\quad
 \includegraphics*[height=0.40\columnwidth]{attack_SCS}}
 \vspace*{.5cm}
\end{figure}

\end{document}